\documentclass[12pt]{article} 
\usepackage{amssymb}
\usepackage{graphicx}
\begin{document} 
\begin{center}
{\large \bf Torus or black disk?}

\vspace{0.5cm}                   

{\bf I.M. Dremin}

\vspace{0.5cm}                       

       P.N. Lebedev Physical Institute RAS, Moscow 119991, Russia\\

\end{center}

\bigskip
UDK 539.12


\bigskip
Tel 89168788639

\bigskip
email dremin@lpi.ru

\bigskip

\begin{abstract}
We show that the interaction region of colliding protons looks completely 
absorptive (black) at the impact parameters up to 0.4 - 0.5 fm at the LHC 
energy 7 TeV. It is governed by the ratio of the elastic diffraction peak slope 
to the total cross section. The corresponding parameter is approximately equal to 1 
at the LHC. The behavior of this ratio at higher energies will show if this region
will evolve to the black disk or to the black torus. Recent fits at 7 TeV
can not distinguish between these possibilities within the limits of 
experimental indefiniteness and of extrapolations in the regions of unmeasured
transferred momenta.
\end{abstract}

The shape of the interaction region of colliding protons changes with
increase of their energies. There were arguments that it looks 
like the completely black disk at asymptotically high energies. Recently
it was shown that this conclusion could be misleading. The black region 
of the size about 0.4 - 0.5 fm is formed at the LHC energy 7 TeV. Its further 
evolution depends on the energy behavior of the ratio of the elastic 
diffraction peak slope to the total cross section. The corresponding parameter 
$Z=4\pi B/\sigma _t$ is approximately equal to 1 at the LHC. Both the total 
cross section of colliding protons $\sigma _t$ and the slope $B$ of the differential 
cross section of elastic scattering increase with energy at high energies. 

For the sake of completeness I have to repeat in the beginning some definitions 
and statements made in my review paper \cite{ufnel} and in my recent paper 
\cite{drjl}.

The differential cross section of elastic scattering $d\sigma /dt$ is 
related to the scattering amplitude $f(s,t)$ in a following way
\begin{equation}
\frac {d\sigma }{dt}=\vert f(s,t)\vert ^2.
\label{dsdt}
\end{equation}
Here $s=4E^2$, where $E$ is the energy in the center of mass system. The 
four-momentum transfer squared is
\begin{equation}
-t=2p^2(1-\cos \theta ) 
\label{trans}
\end{equation}
with $\theta $ denoting the scattering angle in the center of mass system
and $p$ the momentum. The amplitude $f$ is normalized at $t=0$ to the total
cross section by the optical theorem such that
\begin{equation}
{\rm Im}f(s,0)=\sigma _t/\sqrt {16\pi}.
\label{opt}
\end{equation}
Note that the dimension of $f$ is GeV$^{-2}$.

It is known from experiment that protons mostly scatter at rather small
angles within the so-called diffraction cone. As a first approximation, it
can be described by the exponential shape with the slope $B$ such that
\begin{equation}
\frac {d\sigma }{dt}\propto e^{-B\vert t\vert }.
\label{expB}
\end{equation}

To define the geometry of the collision we must express these characteristics
in terms of the transverse distance between the centers of the colliding
protons called the impact parameter $b$. It is easily done by the 
Fourier-Bessel transform of the amplitude $f$ written as
\begin{equation}
i\Gamma (s,b)=\frac {1}{2\sqrt {\pi }}\int _0^{\infty}d\vert t\vert f(s,t)
J_0(b\sqrt {\vert t\vert }).
\label{gamm}
\end{equation}
Using the above formulae, one can write the dimensionless $\Gamma $ as
\begin{equation}
i\Gamma (s,b)=\frac {\sigma _t}{8\pi }\int _0^{\infty}d\vert t\vert 
e^{-B\vert t\vert /2 }(i+\rho (s,t))J_0(b\sqrt {\vert t\vert }).
\label{gam2}
\end{equation}
Here $\rho (s,t) = {\rm Re}f(s,t)/{\rm Im}f(s,t)$ and the diffraction cone
approximation (\ref{expB}) is inserted. Herefrom, one calculates
\begin{equation}
{\rm Re}\Gamma (s,b)=\frac {1}{Z}e^{-\frac {b^2}{2B}},
\label{rega}
\end{equation}
where $Z=4\pi B/\sigma _t$ is the variable used in the review paper 
\cite{ufnel}. This dependence on the impact parameter was used, in particular, 
in \cite{fsw}. 

The elastic scattering amplitude must satisfy the most general principle of
unitarity which states that the total probability of outcomes of any particle 
collision sums to 1 and reads
\begin{equation}
G(s,b)=2{\rm Re}\Gamma (s,b)-\vert \Gamma (s,b)\vert ^2.
\label{unit}
\end{equation}
The left-hand side called the overlap function describes the impact-parameter
profile of inelastic collisions of protons. It satisfies the inequalities
$0\leq G(s,b)\leq 1$ and determines how absoptive is the interaction region
depending on the impact parameter (with $G=1$ for full absorption).

It is known from experiment that the ratio of the real part of the elastic
scattering amplitude to its imaginary part $\rho (s,t)$ is very small at
$t=0$ and, at the beginning, we neglect it and get
\begin{equation}
G(s,b)= \frac {2}{Z}e^{-\frac {b^2}{2B}}-\frac {1}{Z^2}e^{-\frac {b^2}{B}}.
\label{ge}
\end{equation}
For central collisions with $b=0$ it gives
\begin{equation}
G(s,b=0)= \frac {2Z-1}{Z^2}.
\label{gZ}       
\end{equation}
Thus, the darkness of the central region is fully determined by the ratio $Z$. 
It becomes completely absorptive only at $Z=1$ and diminishes for other values
of $Z$. The energy evolution of the parameter $Z$ is shown in the Table 2
of \cite{ufnel}. Here, in the Table, we show the energy evolution of both $Z$
and $G(s,0)$ for $pp$ and $p\bar p$ scattering.
\medskip
\begin{table}
\medskip
Table.  $\;\;$ The energy behavior of $Z$ and $G(s,0)$.
\medskip

    \begin{tabular}{|l|l|l|l|l|l|l|l|l|l|l|l}
        \hline
$\sqrt s$, GeV&2.70&4.11&4.74&7.62&13.8&62.5&546&1800&7000\\ \hline
Z             &0.64&1.02&1.09&1.34&1.45&1.50&1.20&1.08&1.00 \\  
$G(s,0)$     &0.68&1.00&0.993&0.94&0.904&0.89&0.97&0.995&1.00 \\  \hline
   
\end{tabular}
\end{table}

The function $G(s,b)$ in Eq. (\ref{ge}) has the maximum at $b_m^2=-2B\ln Z$
with full absorption $G(b_m)=1$. Its position depends both on $B$ and $Z$.
Note, that, for $Z>1$, one gets $G(s,b)<1$
at any physical $b$ with the largest value reached at $b=0$ because the maximum
appears at non-physical values of $b$. The disk is
semi-transparent. At $Z=1$, the maximum is positioned exactly at $b=0$, and
$G(s,0)=1$. The disk becomes black in the center. At $Z<1$, the maximum shifts
to positive physical impact parameters. The dip is formed at the center. It
becomes deeper at smaller $Z$. The limiting value $Z=0.5$ is considered in
more details below.
\begin{figure}
\centerline{\includegraphics[width=\textwidth, height=9cm]{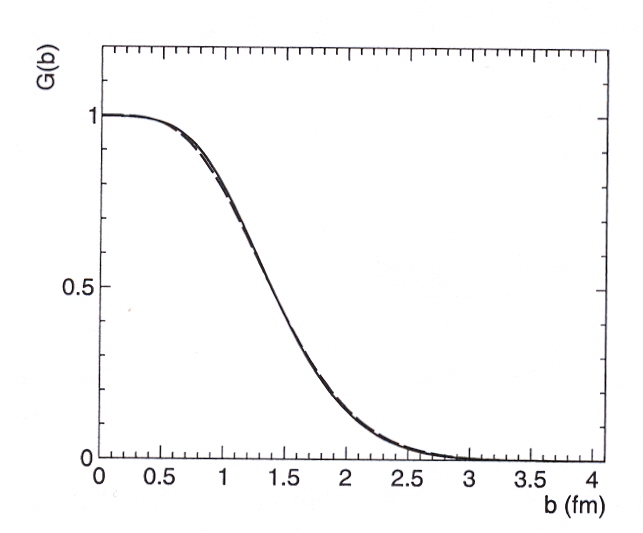}}

Fig. 1 The impact parameter dependence of the overlap function $G(b)$ 

at 7 TeV according to the direct computation from experimental data 

(solid line) and to the diffraction cone approximation (dashed line).

Both curves practically coincide.
\end{figure}

The maximum absorption in central collisions $G(s,0)=1$ is reached at the 
critical point $Z=1$ which is the case (see the table) at $\sqrt s=7$ TeV considered 
first. Moreover, the strongly absorptive core of the interaction region grows in size
as we see from expansion of Eq. (\ref{ge}) at small impact parameters:
\begin{equation}
G(s,b)= \frac {1}{Z^2}[2Z-1-\frac {b^2}{B}(Z-1)-\frac {b^4}{4B^2}(2-Z)].
\label{gb}
\end{equation}
The second term vanishes at $Z=1$, and $G(b)$ develops a plateau which extends
to quite large values of $b$ about 0.4 - 0.5 fm. Even larger values of $b$ 
are necessary for the third term to play any role at 7 TeV where 
$B\approx 20$ GeV$^{-2}$. The structure of the interaction region with a black
central core is also supported by direct computation \cite{dnec} using the 
experimental data of the TOTEM collaboration \cite{totem1, totem2} about
the differential cross section in the region of $\vert t\vert \leq 2.5$ GeV$^2$.
The results of analytical calculations according to Eq. (\ref{ge}) and the numerical 
computation practically coincide (see Fig. 1 borrowed from \cite{ads}). It was also 
shown in \cite{ads} that this two-component structure is well fitted
by the expression with the abrupt (Heaviside-like) change of the exponential.
The diffraction cone contributes mostly to $G(s,b)$. Therefore, the 
large-$\vert t\vert $ elastic scattering can not serve as an effective trigger
of the black core. Inelastic exclusive processes with jets at very high 
multiplicities can be effectively used for this purpose as shown in \cite{ads}. 

It is usually stated that the equality $2Z=8\pi B/\sigma _t=1$ corresponds to 
the black disk limit with equal elastic and inelastic cross sections 
$\sigma _{el}=\sigma _{in}=0.5\sigma _t $. However, one sees from Eq.
(\ref{gZ}) that $G(s,b=0)=0$ in this case, i.e. the interaction region 
is completely transparent in the very central collisions. This paradox is 
resolved if we write the inelastic profile of the interaction region using
Eq. (\ref{ge}). At $Z=0.5$ it looks like
\begin{equation}
G(s,b)= 4[e^{-\frac {b^2}{2B}}-e^{-\frac {b^2}{B}}].
\label{0.5}
\end{equation}
Recalling that $B=R^2/4$, we see that one should rename the black disk as
a black torus (or a black ring)
with full absorption $G(s,b_m)=1$ at the impact parameter 
$b_m=R\sqrt {0.5\ln 2}\approx 0.59R$, complete transparency at $b=0$ and rather
large half-width about 0.7R. Thus, the evolution to values of $Z$ smaller than 
1 at higher energies (if this happens in view of energy tendency of $Z$ shown
in the Table) would imply quite special transition from the two-scale
features at the LHC to torus-like configurations of the interaction region.
Its implications for inelastic processes are to be guessed and studied.
\begin{figure}
\centerline{\includegraphics[width=\textwidth, height=9cm]{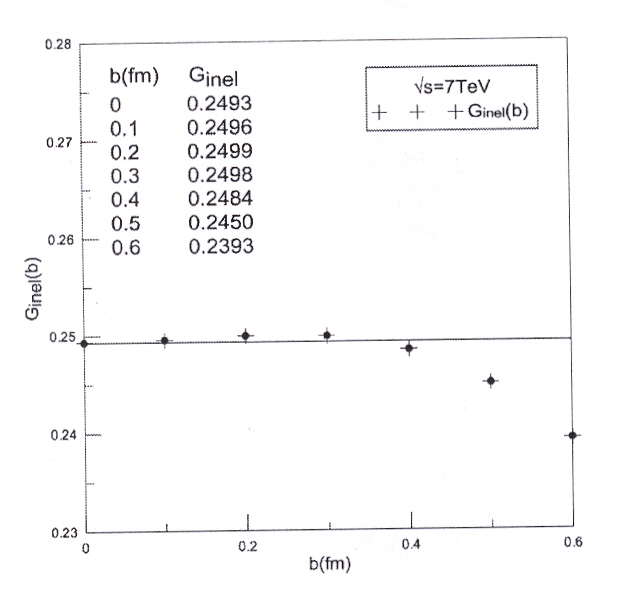}}

Fig. 2 The impact parameter dependence of the function $G_{inel}(b)=0.25 G(b)$

at 7 TeV according to the direct computation from experimental data \cite{mart}.
\end{figure}

It is interesting to note that the authors of the paper \cite{mart} claim that
even at 7 TeV the experimental data show decline from the black disk 
behavior (see Fig. 2 borrowed from \cite{mart}). The transition to the 
black-torus regime in the described above pattern could be noticed in slight 
excess at impact parameters 0.1 - 0.3 fm compared to the very center $b=0$. 
However, it is seen that the excess
is so small that it can be explained first by error bars of experimental data.
There is no such excess in our results \cite{dnec, ads} even though the same 
model was used in both approaches for fits of experimental data. According to 
our computation the last digits look like 97, 96, 96, 95 for $b$ from 0 to 0.3
fm so that there is no excess there but the approximate constancy. The minor  
difference in conclusions attributed to the last digits in the numbers shown
in Fig. 2 for $G_{inel}(b)$ can be ascribed to different procedures adopted 
in these papers for extrapolations to the ranges of transferred momenta 
where there are no experimental data yet. 

Thus it 
seems too early to make any (even preliminary) statements. However, the 
comparison of the results of \cite{dnec, ads} and \cite{mart} shows that we are 
in the critical regime of elastic scattering at 7 TeV as stressed in \cite{drjl}
and should pay special attention to evolution of the parameter $Z$ at higher
energy of 13 TeV which will become available soon.

To conclude, we have shown that the shape of the interaction region of two 
protons colliding at high energies evolves with energy and becomes critical 
at 7 TeV. The absorption at the center of the interaction region of protons 
is determined by a single energy-dependent parameter $Z$. The region of full 
absorption extends to quite large impact parameters if $Z$ tends to 1
that happens at $\sqrt s=7$ TeV. Its difference from 1 at this energy can
not be determined with high enough precision up to now to decide definitely
if the tendency to the new regime has been observed already. Therefore,
the energy behavior of $Z$ at higher energies is especially important in view
of possible evolution of the geometry of the interaction region. 

\medskip

{\bf Acknowledgments}

\medskip 
 
I am grateful for support by the RFBR grants 12-02-91504-CERN-a,
14-02-00099  and the RAS-CERN program.

\end{document}